\documentclass{emulateapj} 
 
\usepackage{verbatim} 
\usepackage{natbib} 
\usepackage{amsmath} 
\usepackage{amsbsy} 
\usepackage{lscape} 
\usepackage{epstopdf} 
 
\slugcomment{Published in ApJL, 682, 9} 
\shorttitle{UV Background and Star Formation} 
\shortauthors{Faucher-Gigu\`ere et al.} 
 
\newcommand{\Lya}{\mbox{Ly$\alpha$}}

\begin{document} 
 
\title{A Flat Photoionization Rate at $2\leq z\leq4.2$: Evidence for a Stellar-Dominated UV Background and Against a Decline of Cosmic Star Formation Beyond $z\sim3$} 
 
\author{Claude-Andr\'e Faucher-Gigu\`ere\altaffilmark{1}, Adam Lidz\altaffilmark{1}, Lars Hernquist\altaffilmark{1}, Matias Zaldarriaga\altaffilmark{1,2}} 
\altaffiltext{1}{Department of Astronomy, Harvard University, Cambridge, MA, 02138, USA; cgiguere@cfa.harvard.edu.} 
\altaffiltext{2}{Jefferson Physical Laboratory, Harvard University, Cambridge, MA, 02138, USA.} 
 
\begin{abstract} 
We investigate the implications of our measurement of the Lyman-$\alpha$ forest opacity at redshifts $2\leq z\leq4.2$ from a sample of 86 high-resolution quasar spectra for the evolution of the cosmic ultraviolet luminosity density and its sources. 
The derived hydrogen photoionization rate $\Gamma$ is remarkably flat over this redshift range, implying an increasing comoving ionizing emissivity with redshift. 
Because the quasar luminosity function is strongly peaked near $z\sim2$, star-forming galaxies likely dominate the ionizing emissivity at $z\gtrsim3$. 
Our measurement 
argues against a star formation rate density declining beyond $z\sim3$, in contrast with existing state-of-the-art determinations of the cosmic star formation history from direct galaxy counts. 
Stellar emission from galaxies therefore likely reionized the Universe. 
\end{abstract} 
 
\keywords{Cosmology: diffuse radiation --- methods: data analysis --- galaxies: formation, evolution, high-redshift --- quasars: absorption lines}  

\section{INTRODUCTION} 
\label{introduction} 
The opacity of the Lyman-$\alpha$ (\Lya) forest is set by a competition between hydrogen photoionizations and recombinations \citep[][]{1965ApJ...142.1633G} and can thus serve as a direct probe of the photonization rate \cite[e.g.][]{1997ApJ...489....7R}. 
The hydrogen photoionization rate $\Gamma$~is a particularly valuable quantity as it is an integral over all sources of ultraviolet (UV) radiation in the Universe, 
\begin{equation} 
\Gamma(z) =  
4\pi \int_{\nu_{\rm HI}}^{\infty}  
\frac{d\nu}{h \nu} 
 J_{\nu}(z) \sigma(\nu), 
\end{equation} 
where $J_{\nu}$ is the angle-averaged specific intensity of the background, $\sigma(\nu)$ is the photoionization cross section of hydrogen, and the integral is from the Lyman limit to infinity. 
As such, it bears a signature of cosmic stellar and quasistellar activity that is not subject to the completeness issues to which direct source counts are prone. 
Moreover, unlike the redshifted radiation backgrounds observed on Earth, the \Lya~forest is a \emph{local} probe of the high-redshift UV radiation, as only sources at approximately the same redshift contribute to $\Gamma$~at any point in the forest \citep[e.g.,][]{1996ApJ...461...20H}. 
In addition to being a powerful probe of galaxy formation and evolution and a fundamental ingredient of cosmological simulations \citep[e.g.,][]{1992MNRAS.256P..43E}, identifying the sources that contribute most to the UV background is key to our understanding of the reionization history of the Universe. 
 
In this \emph{Letter}, we derive the photoionization rate implied by our measurement of the \Lya~forest opacity at $2\leq z\leq4.2$ from a sample of 86 high-resolution quasar spectra \citep[][]{taueffmeas}, for the first time consistently analyzing such a large data set (corrected for both continuum bias and metal absorption) over this redshift interval. 
We discuss the implications of its flatness over this redshift range for the relative contribution of quasars and star-forming galaxies to the high-redshift cosmic UV background. 
Throughout, we assume a \emph{WMAP5} cosmology \citep[][]{2008arXiv0803.0547K}. 
The full details of our analysis, as well as supporting arguments, are presented elsewhere \citep[][]{taueffimp}. 
 
\begin{figure*}[ht] 
\begin{center} 
\includegraphics[width=1\textwidth]{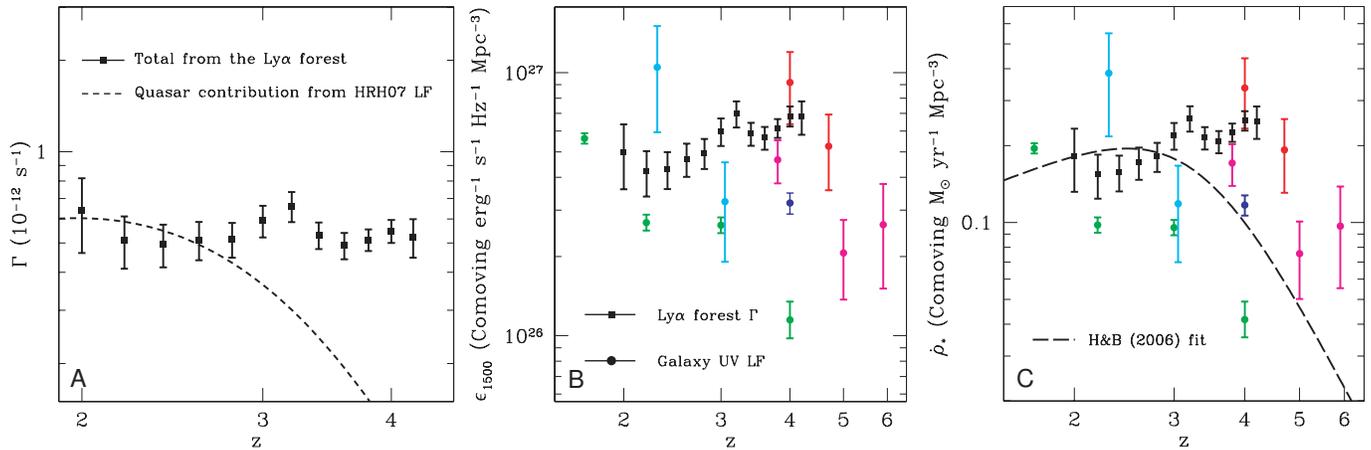} 
\end{center} 
\caption{(A) Photoionization rate $\Gamma$ inferred from our \Lya~effective optical depth measurement (black squares). 
The contribution from quasars calculated using the \cite{2007ApJ...654..731H} quasar luminosity function is shown by the short dashes. 
(B) Comoving UV specific emissivity at 1500~\AA~obtained by integrating galaxy UV luminosity functions. 
The green points are from \cite{2006ApJ...642..653S} (Keck Deep Fields), the cyan points from \cite{2008ApJS..175...48R} (Keck LBG), the red points from \cite{2006ApJ...653..988Y} (Subaru Deep Field), the magenta points from \cite{2007ApJ...670..928B} (Hubble Ultra Deep Field and other deep \emph{HST} fields), and the blue point is based on the \cite{1999ApJ...519....1S} $z\sim4$ LBGs, with the characteristic magnitude and faint-end slope set to the $z\sim3$ values of \cite{2008ApJS..175...48R}. 
See the text for caveats about the error bars shown.
The black points show the emissivity implied by our \Lya~forest measurement, where the normalization was set to match the LF-derived values. 
%minimize the $\chi^{2}$ with respect to LF-derived values. 
(C) Comoving star formation rate density implied by the UV emissivity (see text for details). 
Same color scheme as in (B). 
The long dashed curve shows the best fit of \cite{2006ApJ...651..142H} to the star formation history.} 
\label{gamma etc} 
\end{figure*} 
 
\section{THE PHOTOIONIZATION RATE FROM THE \Lya~FOREST} 
\label{photoionization rate} 
The specific measurement we use is that of the \Lya~effective optical depth $\tau_{\rm eff}$ in $\Delta z=0.2$ bins corrected for continuum bias and for metal absorption following the results of \cite{2003ApJ...596..768S} reported by \cite{taueffmeas}. 
 
The effective optical depth is defined as 
\begin{equation} 
\tau_{\rm eff} \equiv - \ln{[\langle F \rangle(z)]}, 
\end{equation} 
where $\langle F \rangle \equiv \langle \exp{(-\tau)} \rangle$ is the mean transmission of the forest at redshift $z$ and $\tau$ is the local \cite{1965ApJ...142.1633G} optical depth. 
In photoionization equilibrium and for a power-law temperature-density relation for the low-density intergalactic medium (IGM) of the form $T=T_{0}(1+\delta)^{\beta}$ \citep[][]{1997MNRAS.292...27H}, 
\begin{equation} 
\tau = A(z) (1+\delta)^{2-0.7\beta}, 
\end{equation} 
with 
%\begin{equation} 
\begin{align} 
A(z) & \equiv  
\frac{\pi e^{2} f_{\Lya}}{m_{e} \nu_{\Lya}} 
\left( \frac{\rho_{\rm crit} \Omega_{b}}{m_{p}} \right)^{2} 
\frac{1}{H(z)} \notag \\ 
& \times X(X+0.5Y)  
\frac{R_{0} T_{0}^{-0.7}}{\Gamma} 
(1 + z)^{6}. 
\end{align} 
%\end{equation} 
Here, $f_{\Lya}$ is the oscillator strength of the \Lya~transition, $\nu_{\Lya}$ is its frequency, $X$ and $Y$ are the mass fractions of hydrogen and helium \citep[respectively taken to be 0.75 and 0.25;][]{2001ApJ...552L...1B}, $R_{0}=4.2\times10^{-13}~{\rm cm}^{3}~{\rm s}^{-1}/(10^{4}~{\rm K})^{-0.7}$, and $T_{0}$ is the IGM temperature at mean density ($\delta=0$). 
This expression is valid when all the intergalactic helium is fully ionized; an error $\lesssim8\%$ may arise prior to HeII reionization. 
 
Given a volume-weighted probability density function (PDF) for the gas density $\Delta \equiv 1 + \delta$, 
\begin{equation} 
\langle F \rangle(z) = 
\int_{0}^{\infty} 
d\Delta 
P(\Delta; z) 
\exp{(-\tau)}. 
\end{equation} 
We use the analytical fit to gas-dynamical simulations of \cite{2000ApJ...530....1M} for this PDF. 
For the IGM temperature, we interpolate between the $T_{0}\sim2\times10^{4}$ K values measured by \cite{2001ApJ...557..519Z} from the \Lya~forest power spectrum, and $\beta=0.62$, appropriate in the limit of early hydrogen reionization \citep[][]{1997MNRAS.292...27H}. 
 
With the above, we solve for the unique $\Gamma$ that reproduces the measured $\tau_{\rm eff}$ at each redshift. 
The results are shown in Figure \ref{gamma etc}A, where $\Gamma \approx(0.5\pm0.1)\times10^{-12}$ s$^{-1}$ is seen to be remarkably flat over the redshift range $2 \leq z \leq4.2$. 
Note, however, that the absolute normalization of $\Gamma$ depends on the cosmology, the thermal history of the IGM, as well as on the gas density distribution \citep[e.g.,][]{2005MNRAS.357.1178B}, and that a significant scatter between the results of different studies employing the same basic method remains \citep[see Figure 1 of][]{taueffimp}. 
These systematic sources of uncertainty are not included in our analysis. 
On the other hand, the measurement we present consistently samples a large redshift interval with independent statistical errors and should therefore reliably trace the redshift evolution of $\Gamma$. 
 
\section{THE IONIZING SOURCES} 
\label{ionizing sources} 
The UV background is generally assumed to be produced by quasars and star-forming galaxies, but the relative importance of these two populations remains uncertain. 
Moreover, it is unclear whether all the sources responsible for the ionization state of the high-redshift IGM are presently accounted for by magnitude-limited surveys. 
 
In Figure \ref{gamma etc}A, we show the contribution of quasars as calculated using the $B$-band realization of the \cite{2007ApJ...654..731H} bolometric quasar luminosity function (LF). 
The curve, subject to overall normalization uncertainties in the mean free path of ionizing photons, the spectral energy distribution of quasars, and the fraction of ionizing photons that they emit that escape into the IGM, has been renormalized to approximately match the total photoionization rate of the \Lya~forest at $z=2$. 
Its shape is however robustly constrained at redshifts $z\gtrsim2$, owing to both an increasing dominance of the brightest quasars to the UV background and obscuration corrections decreasing in importance with redshift. 
In particular, the quasar contribution to $\Gamma$ is strongly peaked near $z=2$ and even if these objects produce the entire ionizing background at this redshift, they fall short of accounting for the total $\Gamma$ measured at $z=4$ by a factor $\gtrsim5$. 
 
To estimate the contribution of star-forming galaxies to the UV background, we consider recent determinations of the galaxy UV LF from Lyman break galaxy (LBG) surveys by \cite{2006ApJ...642..653S} (Keck Deep Fields), \cite{2007ApJ...670..928B} (Hubble Ultra Deep Field [HUDF] and other deep \emph{Hubble Space Telescope} [HST] deep fields), \cite{1999ApJ...519....1S} and \cite{2008ApJS..175...48R} (Keck LBG), and \cite{2006ApJ...653..988Y} (Subaru Deep Field). 
These were selected to be the most up-to-date measurements in the fields covered. 
The LFs have an effective wavelength near 1500~\AA,~and the specific emissivity at this wavelength is simply obtained by extrapolating and integrating them down to zero luminosity.
The error bars we quote are propagated from those on the individual Schechter parameters; because the latter are generally correlated, these will overestimate the true errors on the luminosity densities.
Exceptions are the \cite{2006ApJ...642..653S} points, for which we take the total luminosity densities and errors reported by the authors.
In order to compare the UV luminosity density of LBGs with our measured $\Gamma$, we convert the latter to a comoving emissivity at 1500~\AA. 
 
Given the proper mean free path of ionizing photons at the Lyman limit $\lambda_{\rm mfp}$, we can calculate the total comoving specific emissivity implied by the measured $\Gamma$: 
\begin{equation} 
\epsilon_{912}\approx 
\frac{h(\alpha_{\rm HI}+3)}{\sigma_{\rm HI} \lambda_{\rm mfp}} (1+z)^{-3} \Gamma 
\end{equation} 
\citep[e.g.,][]{2003ApJ...584..110S} for an ionizing background with a power-law spectrum $J_{\rm HI}\propto \nu^{-\alpha_{\rm HI}}$. 
Here, $\sigma_{\rm HI}$ is the photoionization cross section of hydrogen at the Lyman limit. 
Letting $\alpha_{\rm UV}$ be the spectral index between 912~\AA~and 1500~\AA, we can calculate the emissivity at the wavelength probed by the galaxy UV luminosity functions, 
\begin{equation} 
\epsilon_{1500} =  
\frac{1}{f_{\rm esc}} 
\left( 
\frac{1500~{\rm \AA}}{912~{\rm \AA}} 
\right)^{\alpha_{\rm UV}} 
\epsilon_{912}, 
\end{equation} 
where the escape fraction $f_{\rm esc}$ accounts for the discontinuity at the Lyman limit owing to Lyman-continuum absorption associated with the host galaxy. 
 
The exact value for $f_{\rm esc}$ is not well constrained at present, but is likely to be at most a few percent \citep[e.g.,][]{2001ApJ...546..665S, 2006ApJ...651..688S, 2007ApJ...667L.125C}. 
Here, we simply note that $\epsilon_{1500}\propto \Gamma (1+z)$ for a mean free path $\lambda_{\rm mfp}\propto (1+z)^{4}$, as appropriate if the incidence of Lyman-limit systems increases as $(1+z)^{1.5}$ \citep[][]{1995ApJ...444...64S}. 
This is a conservative assumption, as our conclusions regarding the need for an increasing comoving ionizing emissivity with redshift would only be strengthened if the absorbers responsible for limiting the mean free path instead evolve as $(1+z)^{2}$, as is often assumed on the basis of the better studied lower column density systems \citep[e.g.,][]{1999ApJ...514..648M}. 
We then solve for the normalization that minimizes the $\chi^2$ difference between the emissivities calculated from the UV luminosity functions and our \Lya~forest measurement. 
The result is shown Figure \ref{gamma etc}B. 
In \cite{taueffimp}, we show that for fiducial assumptions, only a small $f_{\rm esc}\sim0.5\%$ is required for LBGs to solely account for the $z\sim3$ ionizing background. 

Within the large scatter, the redshift evolution of UV emissivity derived from the \Lya~forest is reasonably reproduced by the emission from LBGs only. 
The only hint of a decline of the galaxy UV emissivity near $z=4$ comes from the highest-redshift point of \cite{2006ApJ...642..653S}. 
This measurement is inconsistent with the higher points from the Subaru Deep Field \citep[][]{2006ApJ...653..988Y} and \cite{1999ApJ...519....1S}.
This may owe to cosmic variance in the relatively small Keck Deep Fields (169 arcmin$^{2}$ vs. $\sim$850 arcmin$^{2}$ for Subaru and Steidel et al.), or perhaps to selection effects \citep[][]{2008MNRAS.385..493S}.

Quasars being clearly insufficient to solely account for the entire ionizing background implies that galaxies almost certainly dominate at $z\gtrsim3$. 
Measurements of HeII to HI column density ratios however suggest that quasars do contribute a significant, perhaps dominant, fraction of the ionizing background at their $z\sim2$ peak \citep[e.g.,][]{taueffimp}. 
 
\section{THE COSMIC STAR FORMATION HISTORY} 
The above results have interesting implications for the cosmic star formation history. 
In particular, several authors have previously found evidence for a peak in the star formation rate (SFR) density near $z\sim2-3$ \citep[e.g.,][]{1998ApJ...498..106M, 2004ApJ...615..209H, 2006ApJ...651..142H}. 
Barring redshift evolution of dust obscuration, the escape fraction, or the initial mass function (IMF) of stars, the SFR density should trace the UV emissivity, which the \Lya~forest suggests instead increases continuously from $z=2$ to $z=4.2$. 
As a representative example of state-of-the-art determinations of the high-redshift star formation history, we consider the fit of \cite{2006ApJ...651..142H} to a large compilation of galaxy surveys. 
 
In Figure \ref{gamma etc}C, we show the comoving SFR density we derived from the UV emissivities from both the \Lya~forest (assuming that it arises solely from galaxies) and from direct measurements of the galaxy UV LF.  
We convert from specific UV emissivity to SFR density using 
\begin{equation} 
\dot{\rho}_{\star} = 1.08\times10^{-28} \epsilon_{1500}, 
\end{equation} 
where $\dot{\rho}_{\star}$ is in units of comoving M$_{\odot}$ yr$^{-1}$ Mpc$^{-3}$ provided $\epsilon_{\rm 1500}$ is expressed in comoving erg s$^{-1}$ Hz$^{-1}$ Mpc$^{-3}$. 
This conversion is appropriate for a ``modified Salpeter A'' IMF, consistent with the \cite{2006ApJ...651..142H} fit also shown on the Figure. 
Other IMFs would result in different conversion factors. 
However all the data points (and fit) in this plot would be equally renormalized and conclusions with regards to discrepant redshift evolutions would be unaffected. 
We apply a UV obscuration correction factor of 3.4 over the entire redshift range, corresponding to the ``common'' obscuration correction applied by \cite{2004ApJ...615..209H} and \cite{2006ApJ...651..142H} at $z\gtrsim3$. 
Although this correction is unlikely to be exact, it allows for a consistent comparison with the high-redshift fits to the star formation history by these authors. 
 
We find no compelling evidence for a decline in the comoving SFR density over the redshift range probed by our measurement, either from it or from the directly measured UV LF, in contrast to the best fit of \cite{2006ApJ...651..142H}. 
Inspection of Figure \ref{gamma etc}C suggests that the present data are instead roughly consistent with a constant $\dot{\rho}_{\star}\sim0.2$ \mbox{M$_{\odot}$ yr$^{-1}$ Mpc$^{-3}$} at $2\lesssim z \lesssim 4.5$. 
Since our analysis assumes a dust correction consistent with these authors at high redshifts, but is based on more recent data, it thus seems that the SFR density peak suggested by their fit may be an artifact of the scarce high-redshift data in their compilation, which may be affected by cosmic variance and is not uniformly complete. 
For example, one of the $z\sim6$ points that drives the \cite{2006ApJ...651..142H} fit is the estimate of \cite{2004MNRAS.355..374B}, which is only complete to $0.1L^{\star}$ and is based on an extremely small HUDF 11 arcmin$^{2}$ exposure. 
We instead consider the analysis of \cite{2007ApJ...670..928B}, which includes the HUDF data as a subset and yields a higher SFR density, and consistently integrate the LF down to zero luminosity.  
Alternatively, present surveys may be missing a significant UV luminosity density from very faint galaxies. 
 
It is immediately clear from the scatter in panels B and C of Figure \ref{gamma etc} that the total UV luminosity density extrapolated from the measured LF should be interpreted with caution. 
In fact, the dispersion between different points at fixed redshift is generally larger than the calculated error bars, indicating that these are unlikely to be uniformly reliable, a situation which is particularly manifest at $z\sim4$. 
There are several reasons why this may be the case, including extrapolation to fainter magnitudes than probed by individual surveys, cosmic variance arising from large-scale structure, and parameters (perhaps inaccurately) held fixed in some fits. 
 
A number of previous studies of the LBG UV LF have also found little evidence for a decline of the SFR density beyond $z\sim3$ \citep[e.g.,][]{1999ApJ...519....1S, 2004ApJ...600L.103G, 2006ApJ...653..988Y}. 
This finding has in addition been corroborated by measures based on photometric redshifts \citep[e.g.,][]{2001ApJ...546..694T, 2003ApJ...596..748T} and is also in qualitative agreement with theoretical models that predict a SFR history peaking at higher redshift \cite[e.g.,][]{2003MNRAS.339..312S, 2003MNRAS.341.1253H}. 
 
If only because the SFR density is expected to rise continuously on physical grounds, it must eventually decline toward high redshifts. 
\cite{2007ApJ...670..928B} in fact find evidence for such a decline toward $z=6$ on the basis of evolving dust obscuration suggested by observed $\beta$-values at this redshift \citep[e.g.,][]{2005MNRAS.359.1184S}. 
We simply contend here that neither the present \Lya~forest data or the recent UV LF compiled here, especially when considered together with their mutual scatter after extrapolation down to zero luminosity, show convincing evidence for the often-assumed peak in SFR density near $z\sim2-3$. 
The requirement that the Universe be reionized by $z=6$ also supports a SFR peaking significantly earlier \citep[][]{taueffimp}. 
 
\section{COMPARISON WITH PREVIOUS WORK} 
Similar conclusions have been reached in previous studies of the UV background. 
\cite{2005MNRAS.357.1178B}, in particular, inferred $\Gamma$ from the \Lya~opacity measurement of \cite{2003ApJ...596..768S} and also found its evolution to be consistent with being constant at $2\leq z\leq4$. 
By comparing with the estimated quasar contribution, they also found evidence for a stellar-dominated UV background at all redshifts. 
In their analysis using the \cite{2000ApJ...530....1M} PDF, \cite{2007ApJ...662...72B} also derived a flat $\Gamma$ over this redshift range. 
Our results extend beyond previous analyses in highlighting that common assumptions regarding the star formation history fall short of providing for the ionizing rate of the forest at $z\gtrsim3$. 
 
Measurements based on the proximity effect \citep[e.g.,][]{2000ApJS..130...67S} have tended to yield $\Gamma$ values higher by a factor of $\sim3$. 
However, the overdense regions in which quasars reside are likely to bias these measurements high \cite[][]{2008ApJ...673...39F}. 
 
\section{REIONIZATION} 
\label{reionization} 
The decline of the quasar LF and the increasing dominance of stellar emission to the high-redshift $z\gtrsim3$ ionizing background make a compelling case that the Universe was reionized by stars. 
This gives credibility to analytical and numerical calculations of hydrogen reionization that make this assumption \citep[e.g.,][]{2004ApJ...613....1F, 2007MNRAS.377.1043M, 2007ApJ...654...12Z}. 
This is encouraging news for upcoming observational probes of the epoch of reionization, such as redshifted 21-cm emission and high-redshift \Lya~emitters \citep[e.g.,][]{2004ApJ...608..622Z,2007MNRAS.381...75M}, whose detailed interpretation will rely on our understanding of the morphology of reionization and its origin. 
 
\acknowledgements 
We thank Rychard Bouwens, Andrew Hopkins, Philip Hopkins, Matthew McQuinn, and Jason X. Prochaska for useful discussions. 
CAFG is supported by a NSERC Postgraduate Fellowship and the Canadian Space Agency. 
This work was supported in part by NSF grants ACI 96-19019, AST 00-71019, AST 02-06299, AST 03-07690, and AST 05-06556, and NASA ATP grants NAG5-12140, NAG5-13292, NAG5-13381, and NNG-05GJ40G. 
Further support was provided by the David and Lucile Packard, the Alfred P. Sloan, and the John D. and Catherine T. MacArthur Foundations. 
 
\bibliography{references} 

\begin{thebibliography}{44}
\expandafter\ifx\csname natexlab\endcsname\relax\def\natexlab#1{#1}\fi

\bibitem[{{Becker} {et~al.}(2007){Becker}, {Rauch}, \&
  {Sargent}}]{2007ApJ...662...72B}
{Becker}, G.~D., {Rauch}, M., \& {Sargent}, W.~L.~W. 2007, \apj, 662, 72

\bibitem[{{Bolton} {et~al.}(2005){Bolton}, {Haehnelt}, {Viel}, \&
  {Springel}}]{2005MNRAS.357.1178B}
{Bolton}, J.~S., {Haehnelt}, M.~G., {Viel}, M., \& {Springel}, V. 2005, \mnras,
  357, 1178

\bibitem[{{Bouwens} {et~al.}(2007){Bouwens}, {Illingworth}, {Franx}, \&
  {Ford}}]{2007ApJ...670..928B}
{Bouwens}, R.~J., {Illingworth}, G.~D., {Franx}, M., \& {Ford}, H. 2007, \apj,
  670, 928

\bibitem[{{Bunker} {et~al.}(2004){Bunker}, {Stanway}, {Ellis}, \&
  {McMahon}}]{2004MNRAS.355..374B}
{Bunker}, A.~J., {Stanway}, E.~R., {Ellis}, R.~S., \& {McMahon}, R.~G. 2004,
  \mnras, 355, 374

\bibitem[{{Burles} {et~al.}(2001){Burles}, {Nollett}, \&
  {Turner}}]{2001ApJ...552L...1B}
{Burles}, S., {Nollett}, K.~M., \& {Turner}, M.~S. 2001, \apjl, 552, L1

\bibitem[{{Chen} {et~al.}(2007){Chen}, {Prochaska}, \&
  {Gnedin}}]{2007ApJ...667L.125C}
{Chen}, H.-W., {Prochaska}, J.~X., \& {Gnedin}, N.~Y. 2007, \apjl, 667, L125

\bibitem[{{Efstathiou}(1992)}]{1992MNRAS.256P..43E}
{Efstathiou}, G. 1992, \mnras, 256, 43P

\bibitem[{{Faucher-Gigu{\`e}re}
  {et~al.}(2008{\natexlab{a}}){Faucher-Gigu{\`e}re}, {Lidz}, {Hernquist}, \&
  {Zaldarriaga}}]{taueffimp}
{Faucher-Gigu{\`e}re}, C.~A., {Lidz}, A., {Hernquist}, L., \& {Zaldarriaga}, M.
  2008{\natexlab{a}}, \apj, in prep.

\bibitem[{{Faucher-Gigu{\`e}re}
  {et~al.}(2008{\natexlab{b}}){Faucher-Gigu{\`e}re}, {Lidz}, {Zaldarriaga}, \&
  {Hernquist}}]{2008ApJ...673...39F}
{Faucher-Gigu{\`e}re}, C.-A., {Lidz}, A., {Zaldarriaga}, M., \& {Hernquist}, L.
  2008{\natexlab{b}}, \apj, 673, 39

\bibitem[{{Faucher-Gigu{\`e}re}
  {et~al.}(2008{\natexlab{c}}){Faucher-Gigu{\`e}re}, {Prochaska}, {Lidz},
  {Hernquist}, \& {Zaldarriaga}}]{taueffmeas}
{Faucher-Gigu{\`e}re}, C.~A., {Prochaska}, J.~X., {Lidz}, A., {Hernquist}, L.,
  \& {Zaldarriaga}, M. 2008{\natexlab{c}}, \apj, in press

\bibitem[{{Furlanetto} {et~al.}(2004){Furlanetto}, {Zaldarriaga}, \&
  {Hernquist}}]{2004ApJ...613....1F}
{Furlanetto}, S.~R., {Zaldarriaga}, M., \& {Hernquist}, L. 2004, \apj, 613, 1

\bibitem[{{Giavalisco} {et~al.}(2004){Giavalisco}, {Dickinson}, {Ferguson}, \&
  {...}}]{2004ApJ...600L.103G}
{Giavalisco}, M., {Dickinson}, M., {Ferguson}, H.~C., \& {...} 2004, \apjl,
  600, L103

\bibitem[{{Gunn} \& {Peterson}(1965)}]{1965ApJ...142.1633G}
{Gunn}, J.~E., \& {Peterson}, B.~A. 1965, \apj, 142, 1633

\bibitem[{{Haardt} \& {Madau}(1996)}]{1996ApJ...461...20H}
{Haardt}, F., \& {Madau}, P. 1996, \apj, 461, 20

\bibitem[{{Hernquist} \& {Springel}(2003)}]{2003MNRAS.341.1253H}
{Hernquist}, L., \& {Springel}, V. 2003, \mnras, 341, 1253

\bibitem[{{Hopkins}(2004)}]{2004ApJ...615..209H}
{Hopkins}, A.~M. 2004, \apj, 615, 209

\bibitem[{{Hopkins} \& {Beacom}(2006)}]{2006ApJ...651..142H}
{Hopkins}, A.~M., \& {Beacom}, J.~F. 2006, \apj, 651, 142

\bibitem[{{Hopkins} {et~al.}(2007){Hopkins}, {Richards}, \&
  {Hernquist}}]{2007ApJ...654..731H}
{Hopkins}, P.~F., {Richards}, G.~T., \& {Hernquist}, L. 2007, \apj, 654, 731

\bibitem[{{Hui} \& {Gnedin}(1997)}]{1997MNRAS.292...27H}
{Hui}, L., \& {Gnedin}, N.~Y. 1997, \mnras, 292, 27

\bibitem[{{Komatsu} {et~al.}(2008){Komatsu}, {Dunkley}, {Nolta}, {Bennett},
  {Gold}, {Hinshaw}, {Jarosik}, {Larson}, {Limon}, {Page}, {Spergel},
  {Halpern}, {Hill}, {Kogut}, {Meyer}, {Tucker}, {Weiland}, {Wollack}, \&
  {Wright}}]{2008arXiv0803.0547K}
{Komatsu}, E., {Dunkley}, J., {Nolta}, M.~R., {Bennett}, C.~L., {Gold}, B.,
  {Hinshaw}, G., {Jarosik}, N., {Larson}, D., {Limon}, M., {Page}, L.,
  {Spergel}, D.~N., {Halpern}, M., {Hill}, R.~S., {Kogut}, A., {Meyer}, S.~S.,
  {Tucker}, G.~S., {Weiland}, J.~L., {Wollack}, E., \& {Wright}, E.~L. 2008,
  ArXiv e-prints, 803

\bibitem[{{Madau} {et~al.}(1999){Madau}, {Haardt}, \&
  {Rees}}]{1999ApJ...514..648M}
{Madau}, P., {Haardt}, F., \& {Rees}, M.~J. 1999, \apj, 514, 648

\bibitem[{{Madau} {et~al.}(1998){Madau}, {Pozzetti}, \&
  {Dickinson}}]{1998ApJ...498..106M}
{Madau}, P., {Pozzetti}, L., \& {Dickinson}, M. 1998, \apj, 498, 106

\bibitem[{{McQuinn} {et~al.}(2007{\natexlab{a}}){McQuinn}, {Hernquist},
  {Zaldarriaga}, \& {Dutta}}]{2007MNRAS.381...75M}
{McQuinn}, M., {Hernquist}, L., {Zaldarriaga}, M., \& {Dutta}, S.
  2007{\natexlab{a}}, \mnras, 381, 75

\bibitem[{{McQuinn} {et~al.}(2007{\natexlab{b}}){McQuinn}, {Lidz}, {Zahn},
  {Dutta}, {Hernquist}, \& {Zaldarriaga}}]{2007MNRAS.377.1043M}
{McQuinn}, M., {Lidz}, A., {Zahn}, O., {Dutta}, S., {Hernquist}, L., \&
  {Zaldarriaga}, M. 2007{\natexlab{b}}, \mnras, 377, 1043

\bibitem[{{Miralda-Escud{\'e}} {et~al.}(2000){Miralda-Escud{\'e}}, {Haehnelt},
  \& {Rees}}]{2000ApJ...530....1M}
{Miralda-Escud{\'e}}, J., {Haehnelt}, M., \& {Rees}, M.~J. 2000, \apj, 530, 1

\bibitem[{{Rauch} {et~al.}(1997){Rauch}, {Miralda-Escude}, {Sargent}, \&
  {...}}]{1997ApJ...489....7R}
{Rauch}, M., {Miralda-Escude}, J., {Sargent}, W.~L.~W., \& {...} 1997, \apj,
  489, 7

\bibitem[{{Reddy} {et~al.}(2008){Reddy}, {Steidel}, {Pettini}, {Adelberger},
  {Shapley}, {Erb}, \& {Dickinson}}]{2008ApJS..175...48R}
{Reddy}, N.~A., {Steidel}, C.~C., {Pettini}, M., {Adelberger}, K.~L.,
  {Shapley}, A.~E., {Erb}, D.~K., \& {Dickinson}, M. 2008, \apjs, 175, 48

\bibitem[{{Sawicki} \& {Thompson}(2006)}]{2006ApJ...642..653S}
{Sawicki}, M., \& {Thompson}, D. 2006, \apj, 642, 653

\bibitem[{{Schaye} {et~al.}(2003){Schaye}, {Aguirre}, {Kim}, {Theuns}, {Rauch},
  \& {Sargent}}]{2003ApJ...596..768S}
{Schaye}, J., {Aguirre}, A., {Kim}, T.-S., {Theuns}, T., {Rauch}, M., \&
  {Sargent}, W.~L.~W. 2003, \apj, 596, 768

\bibitem[{{Schirber} \& {Bullock}(2003)}]{2003ApJ...584..110S}
{Schirber}, M., \& {Bullock}, J.~S. 2003, \apj, 584, 110

\bibitem[{{Scott} {et~al.}(2000){Scott}, {Bechtold}, {Dobrzycki}, \&
  {Kulkarni}}]{2000ApJS..130...67S}
{Scott}, J., {Bechtold}, J., {Dobrzycki}, A., \& {Kulkarni}, V.~P. 2000, \apjs,
  130, 67

\bibitem[{{Shapley} {et~al.}(2006){Shapley}, {Steidel}, {Pettini},
  {Adelberger}, \& {Erb}}]{2006ApJ...651..688S}
{Shapley}, A.~E., {Steidel}, C.~C., {Pettini}, M., {Adelberger}, K.~L., \&
  {Erb}, D.~K. 2006, \apj, 651, 688

\bibitem[{{Springel} \& {Hernquist}(2003)}]{2003MNRAS.339..312S}
{Springel}, V., \& {Hernquist}, L. 2003, \mnras, 339, 312

\bibitem[{{Stanway} {et~al.}(2008){Stanway}, {Bremer}, \&
  {Lehnert}}]{2008MNRAS.385..493S}
{Stanway}, E.~R., {Bremer}, M.~N., \& {Lehnert}, M.~D. 2008, \mnras, 385, 493

\bibitem[{{Stanway} {et~al.}(2005){Stanway}, {McMahon}, \&
  {Bunker}}]{2005MNRAS.359.1184S}
{Stanway}, E.~R., {McMahon}, R.~G., \& {Bunker}, A.~J. 2005, \mnras, 359, 1184

\bibitem[{{Steidel} {et~al.}(1999){Steidel}, {Adelberger}, {Giavalisco},
  {Dickinson}, \& {Pettini}}]{1999ApJ...519....1S}
{Steidel}, C.~C., {Adelberger}, K.~L., {Giavalisco}, M., {Dickinson}, M., \&
  {Pettini}, M. 1999, \apj, 519, 1

\bibitem[{{Steidel} {et~al.}(2001){Steidel}, {Pettini}, \&
  {Adelberger}}]{2001ApJ...546..665S}
{Steidel}, C.~C., {Pettini}, M., \& {Adelberger}, K.~L. 2001, \apj, 546, 665

\bibitem[{{Stengler-Larrea} {et~al.}(1995){Stengler-Larrea}, {Boksenberg},
  {Steidel}, \& {...}}]{1995ApJ...444...64S}
{Stengler-Larrea}, E.~A., {Boksenberg}, A., {Steidel}, C.~C., \& {...} 1995,
  \apj, 444, 64

\bibitem[{{Thompson}(2003)}]{2003ApJ...596..748T}
{Thompson}, R.~I. 2003, \apj, 596, 748

\bibitem[{{Thompson} {et~al.}(2001){Thompson}, {Weymann}, \&
  {Storrie-Lombardi}}]{2001ApJ...546..694T}
{Thompson}, R.~I., {Weymann}, R.~J., \& {Storrie-Lombardi}, L.~J. 2001, \apj,
  546, 694

\bibitem[{{Yoshida} {et~al.}(2006){Yoshida}, {Shimasaku}, {Kashikawa}, \&
  {...}}]{2006ApJ...653..988Y}
{Yoshida}, M., {Shimasaku}, K., {Kashikawa}, N., \& {...} 2006, \apj, 653, 988

\bibitem[{{Zahn} {et~al.}(2007){Zahn}, {Lidz}, {McQuinn}, {Dutta}, {Hernquist},
  {Zaldarriaga}, \& {Furlanetto}}]{2007ApJ...654...12Z}
{Zahn}, O., {Lidz}, A., {McQuinn}, M., {Dutta}, S., {Hernquist}, L.,
  {Zaldarriaga}, M., \& {Furlanetto}, S.~R. 2007, \apj, 654, 12

\bibitem[{{Zaldarriaga} {et~al.}(2004){Zaldarriaga}, {Furlanetto}, \&
  {Hernquist}}]{2004ApJ...608..622Z}
{Zaldarriaga}, M., {Furlanetto}, S.~R., \& {Hernquist}, L. 2004, \apj, 608, 622

\bibitem[{{Zaldarriaga} {et~al.}(2001){Zaldarriaga}, {Hui}, \&
  {Tegmark}}]{2001ApJ...557..519Z}
{Zaldarriaga}, M., {Hui}, L., \& {Tegmark}, M. 2001, \apj, 557, 519

\end{thebibliography}
 
\end{document}